\documentclass{aa}

\usepackage{graphicx}
\usepackage{natbib}
\usepackage{amsmath}

\def\tr{\textrm}

\def\Msol{\tr{M}_{\odot}{}}

\def\Lb{L_{\tty{bol}}}
\def\Lx{L_{[0.5-2\,\tss{keV}]}}
\def\Lhx{L_{[2-10\,\tss{keV}]}}

\newcommand{\tty}[1]{\textrm{\tiny #1}}
\newcommand{\tss}[1]{\textrm{\scriptsize #1}}
\newcommand{\unit}[2]{\tr{#1}^{#2}}

\def\be{\begin{eqnarray}}                   
\def\ee{\end{eqnarray}}                     
\newcommand{\eqp}[1]{(\ref{#1})}                   
\newcommand{\eqpd}[2]{(\ref{#1}) and (\ref{#2})}    
\newcommand{\equ}[1]{(see Eq.~\ref{#1})}
\newcommand{\tab}[1]{(see Table~\ref{#1})}

\begin{document}

\title{Ratio of energies radiated in the universe through accretive processes and nucleosynthesis}

\author{Juan~A. Zurita Heras\inst{1}\fnmsep\inst{2},
	Marc T\"urler\inst{1}\fnmsep\inst{2} and
	Thierry~J.-L. Courvoisier\inst{1}\fnmsep\inst{2}}

\offprints{J.A. Zurita Heras \\
	  \email{Juan.Zurita@obs.unige.ch}}

\institute{INTEGRAL Science Data Centre, ch. d'Ecogia 16, 1290 Versoix, Switzerland
	  \and Observatoire de Gen\`eve, ch. des Maillettes 51, 1290 Sauverny, Switzerland}

\date{Received / Accepted }

\abstract{
	 We present here a new determination of the ratio of energies radiated
	 by active galactic nuclei and by stars and discuss the reasons for the
	 apparently conflicting results found in previous studies.
	 We conclude that the energy radiated by accretion processes onto super
	 massive black holes is about 1 to 5\% of the energy radiated by stars.
 	 We also estimate that the total mass accreted on average by a super
 	 massive black hole at the centre of a typical $10^{11}\,\Msol$ galaxy
 	 is of about $7\,10^{7}\,\Msol$.
	 }

\authorrunning{Zurita et al.}
\maketitle

\section{Introduction}\label{intro}

The universe is composed of various objects with a wide variety of emission
processes. However, photons have three main origins: the big bang as the source
of the cosmic microwave background (CMB), nuclear reactions in stars and gravity
through accretion onto compact objects, especially super massive black holes
(SMBH). Currently (at $z\simeq 0$), the ratio of the number of photons of CMB
and stellar origin in the universe is of the order of 400. It is more difficult
to estimate the ratio of the nuclear energy radiated by stars and the
gravitational energy radiated by active galactic nuclei (AGN). Several recent
studies have shown that the energy released by nucleosynthesis is evolving with
cosmic time \citep{Madaual96,Madaual98,Trenthamal99,Somervilleal01}. It further
seems that the star formation history roughly matches the evolution of energy
release through accretion onto black holes \citep{Dunlop98,Franceschinial99},
since both processes peak around $z\sim 2$. It is therefore possible to relate
both phenomena and to estimate the energy ratio radiated by AGN and by stars
over the history of the universe.

The determination of this ratio has been addressed several times in the
literature. \citet{Dunlop98} constructed two models of the radio luminosity
function of a sample of radio-loud quasars (RLQ), the first only considering
luminosity evolution and the second combining luminosity and density evolution.
He then related this luminosity to the mass accreted onto the central black hole
and compared these curves with the star formation history. The two curves appear
to be correlated, suggesting that when 1 $\Msol$ is accreted by a SMBH in a
radio loud quasar, $10^{7}\,\Msol$ are used in the star formation process. Based
on this relation, \citet{Csier01} derived a ratio of the energy radiated by RLQ
and stars of $E_{\tty{RLQ}}$/$E_{\tty{stars}}\sim 10^{-5}$ by assuming an
accretion efficiency of 10\,\% for the RLQ and a stellar population made of 10
solar mass stars each radiating $4\,10^{52}$ ergs over their lifetime. Rather
than considering only radio loud AGN, \citet{Franceschinial99} use the $0.5-2$
keV X-ray emission as a measure of the energy radiated by AGN. They find that
when 1 $\Msol\,\unit{yr}{-1}\,\unit{Mpc}{-3}$ is absorbed for star formation,
the $0.5-2$ keV volume emissivity from AGN is $2.4\,10^{40}\
\tr{ergs}\,\unit{s}{-1}\,\unit{Mpc}{-3}$. Considering that type I and II AGN
bolometric luminosity is 250 times the $0.5-2$ keV luminosity and a stellar
radiative efficiency of 0.001, \citet{Franceschinial99} finally obtain
$\Lb\tss{(AGN)}$/$\Lb\tss{(SFR)}\sim 0.1$. 

A different approach was followed by \citet{FabIwa99}. Their estimate of this
ratio is based on relations linking the bulge mass of a galaxy to both the mass
of its central black hole and of its stars. They obtained a value of $\sim$\,0.2
for the ratio of the energy radiated by accretion processes and by stars, when
assuming an accretion efficiency of 10\,\% for the AGN and the fact that one
tenth of the stellar mass is used for nuclear fusion with an efficiency of
0.6\,\%.

Finally, \citet{Elvisal02} estimated that AGN contribute by at least 7\,\% to
the total luminosity of the universe as derived from the diffuse background from
submillimeter to ultraviolet wavelengths. They obtained this result by first
deriving a lower limit of the AGN X-ray emission from the X-ray background (XRB)
and by applying a bolometric correction determined with the average spectral
energy distribution of quasars.

The aim of this study is to compare the various studies described above and to
derive a new estimate of the ratio of gravitational energy released around SMBH
to the energy released by nuclear fusion in stars. We first derive the radiation
energy density due to stars in Sect.~\ref{EnStars}, then the corresponding value
for accretion by SMBH based on the XRB in Sect.~\ref{EnAGNs}. The obtained ratio
is compared with previous studies in Sect.~\ref{discussion}, where we try to
identify the origin of the discrepancies.

\section{Energy radiated by stars}\label{EnStars}

To estimate the energy radiated by stars in the universe we need to know both
the energy release of a typical stellar population and the overall star
formation history. Below, we start with the calculation of the stellar energy
release, while the evolution of the star density will be described in
Sect.~\ref{evolStars}.

\subsection{Stellar energy release}

We use the \textit{Starburst
99}\footnote{http://www.stsci.edu/science/starburst99/} models of
\citet{Leithereral99} to determine the typical stellar energy release. These
models predict the spectrophotometric evolution of starburst galaxies between
$10^{6}$ and $10^{9}$ years after the onset of star formation based on the
stellar evolution models of the Geneva group
\citep{Schalleral92,Charbonnelal93,Schaereral93a,Schaereral93b,Meynetal94}. 
They consider the atmosphere models of \citet{Lejeuneal97} and those of
\citet{Schmutzal92} when the mass loss becomes important. A simple black body is
used for cool stars with additional nebular continuum including free-free
interactions below 912\,\AA\ and free-bound interactions above. These models
have been computed with the isochrone synthesis method and are optimized for
massive stars.

\begin{figure}
 \resizebox{\hsize}{!}{\includegraphics{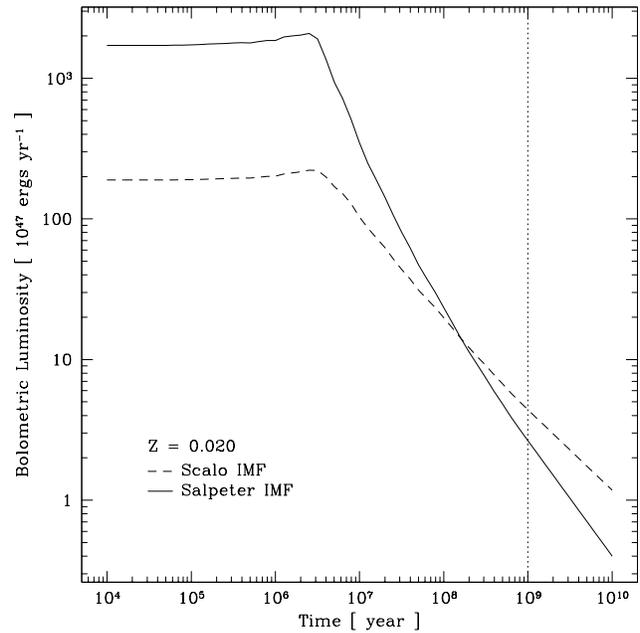}}
 \caption{
	  Evolution of the bolometric luminosity of a $10^6\,\Msol$ star cluster
	  formed instantaneously with a metallicity $Z=0.020$ according to the
	  model of \citet{Leithereral99}. The solid line is for a Salpeter IMF
	  ($\alpha\!=\!2.35$) and the dashed line is for a Scalo IMF
	  ($\alpha\!=\!3.3$). Only stars in the mass range between 1 and
	  $100\,\Msol$ are considered. Beyond $10^{9}$ years (dotted line), the
	  model data have been linearly extrapolated up to $10^{10}$ years.
	 }
 \label{LumI020}
\end{figure}

Since we are only interested in the total energy release of a typical
population of stars during its whole life, we only consider the instantaneous
star formation models of \citet{Leithereral99} because in this case most of the
energy is released before $10^{9}$ years. To assess the effect of changing the
powerlaw index $\alpha$ of the initial mass function (IMF) of the stars we
consider both a Salpeter IMF ($\alpha\!=\!2.35$) and a steeper Scalo IMF
($\alpha\!=\!3.3$). In both cases, the cutoff masses are chosen as
$M_{\tty{low}}\!=\!1\,\Msol$ and $M_{\tty{up}}\!=\!100\,\Msol$. The effect of
changing the metallicity $Z$ is taken into account by considering four different
metallicities: $Z=0.040, 0.020 (=Z_{\odot}), 0.008$ and $0.001$, but without
chemical evolution in the models. As an example, we show in Fig.~\ref{LumI020}
the evolution of the bolometric luminosity of a $10^6\,\Msol$ star cluster
formed instantaneously with a solar metallicity ($Z=0.020$) according to the
model of \citet{Leithereral99}. We extrapolate the bolometric luminosity from
$10^{9}$ to $10^{10}$ years with a power-law in order to include the energy
radiated during the final stages of stellar activity. This extrapolation is in
good agreement with the earlier study of \citet{ChaBru91}.

The total energy radiated by a $10^6\,\Msol$ star cluster is obtained by
integrating the bolometric luminosity over time.
\begin{table}
 \begin{center}
 \caption{\label{tab_star} 
	  Energy radiated per unit of solar mass by stars
	  $(E_{\tty{rad}}/\Msol)_{\tty{stars}}$ and the corresponding energy
	  density $U_{\tty{stars}}$ for different metallicities $Z$ and for a
	  Salpeter ($\alpha\!=\!2.35$) or a Scalo ($\alpha\!=\!3.3$) IMF. The
	  last column shows the corresponding energy density ratio radiated by
	  AGN and by stars. 
	 }
 \begin{tabular}{ccccc}
  \hline
  \hline
  Z & $\alpha_{\tty{IMF}}$ & $(E_{\tty{rad}}/\Msol)_{\tty{stars}}$ & $U_{\tty{stars}}$ & $U_{\tty{AGN}}$/$U_{\tty{stars}}$ \\
   & & \scriptsize{$10^{51}\ \tr{ergs}\,\Msol^{-1}$} & \tiny{$10^{60}\ \tr{ergs}\,\unit{Mpc}{-3}$} & \tiny$(10^{-2})$ \\ 
  \hline
  $0.001$ & 2.35 & 4.16 & 1.29 & 2.8 \\
  	  & 3.3  & 4.69 & 3.33 & 1.1 \\[1mm]
  $0.008$ & 2.35 & 3.46 & 1.07 & 3.4 \\
 	  & 3.3  & 3.39 & 2.40 & 1.5 \\[1mm]
  $0.020$ & 2.35 & 3.09 & 0.96 & 3.8 \\
 	  & 3.3  & 2.92 & 2.07 & 1.8 \\[1mm]
  $0.040$ & 2.35 & 2.33 & 0.72 & 5.1 \\
 	  & 3.3  & 1.71 & 1.21 & 3.0 \\
  \hline
 \end{tabular}
 \end{center}
\end{table}
The results are presented in Table~\ref{tab_star} for the instantaneous stellar
formation law, four different metallicities, and two different IMF.
Since the bolometric luminosity is dominated by the energy radiated by the most
massive stars, the flatter powerlaw index $\alpha$ of the Salpeter IMF provides
a higher energy release than the Scalo IMF until $\sim 10^{8}$ years.
Afterwards, the radiated energy being provided by less massive stars, a Scalo
IMF gives more energy. Globally, the difference between the total energy
radiated by a cluster of stars for a Scalo or a Salpeter IMF is quite small. On
the other hand, the effect of increasing the metallicity is to decrease the
stellar energy release. This effect can be understood as being due to an
increase of stellar opacity with metallicity \citep{Mowlavial98}. A quick
comparison with \citet{Csier01} shows that the values in Table~\ref{tab_star}
are consistent with his rough estimation that stars radiate $\sim 10^{51}\
\tr{ergs}\,\tr{M}_{\odot}^{-1}$ over their entire lives.

\subsection{Evolution of the stellar density}\label{evolStars}

\begin{figure}
 \resizebox{\hsize}{!}{\includegraphics{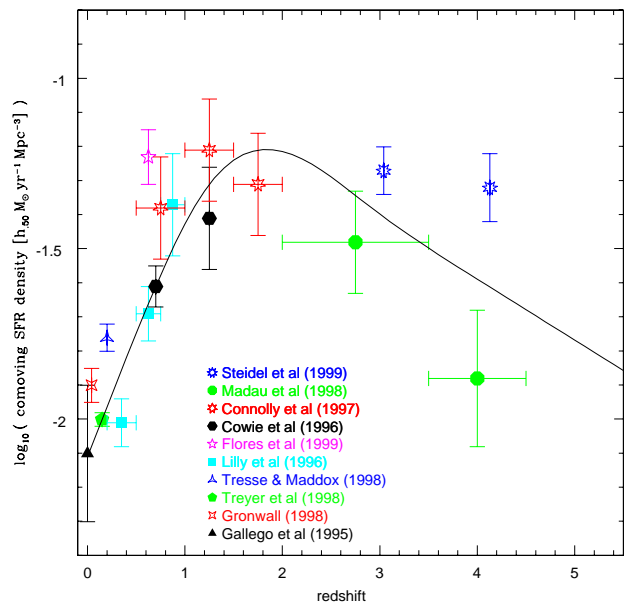}}
 \caption{
	  Evolution of the comoving SFR density for an Einstein-de Sitter
	  cosmology and a Salpeter IMF from the data in Table A2 of
	  \citet{Somervilleal01}. The solid line is a cubic spline matching as
	  well as possible the observational constraints.
	 }
 \label{SFRdensity}
\end{figure}

\citet{Madaual96} first computed the star formation as a function of redshift.
Since then, several authors have added new points to his diagram of the star
formation rate (SFR) per unit of comoving volume as a function of the redshift.
Different tracers have been used to derive the SFR, but now, it is common to
use galaxy luminosities at different wavelengths. They can be converted into
stellar formation rates using stellar population and galaxy spectral models,
stellar formation scenarios and various IMF. \citet{Madaual98} propose
conversion factors from luminosity to star formation rates ($L_{\tty{UV}}=C\cdot
SFR$) with different values of $C$ for a Salpeter IMF and a Scalo IMF.
\citet{Somervilleal01} compiled all the observations made in this way and
present a homogeneous table of the comoving SFR density data for different
cosmological models and a Salpeter IMF. Fig.~\ref{SFRdensity} shows those data
in the case of an Einstein-de Sitter cosmology and a Salpeter IMF.
We calculate the stellar density $\rho_{\tty{stars}}$ by integrating the star
formation rate $\dot{\rho}_{\tty{stars}}$ over the whole cosmic time:
\be
\rho_{\tty{stars}}=\int{\dot{\rho}_{\tty{stars}}\,dt}\label{rho*}\,.
\ee
If we consider a Friedman cosmological model, a flat universe, $q_{0}=0.5$, and
$H_{0}=50\ \tr{km}\,\unit{s}{-1}\,\unit{Mpc}{-1}$, Eq.~(\ref{rho*}) becomes:
\be
\rho_{\tty{stars}}=\int_{z_{\tty{max}}}^{0}{\dot{\rho}_{\tty{stars}}(z)\frac{dz}{H_{0}(1+z)^{2.5}}}\,.
\ee
With $z_{\tty{max}}=5.5$ and $\dot{\rho}_{\tty{stars}}(z)$ being the solid line
adjusted to the data points of Fig.~\ref{SFRdensity}, we obtain:
\be
\rho_{\tty{stars}}=3.1\cdot 10^{8}\ \Msol\,\unit{Mpc}{-3}\,.
\label{rho_salp_eds}
\ee
This value is an average over the star formation history. Using the conversion
factors given by \citet{Madaual98} and the table of \citet{Somervilleal01}, we
deduce a different stellar density if we consider a Scalo IMF. Thus, for a Scalo
IMF, an Einstein-de Sitter cosmology and
$H_{0}=50\ \tr{km}\,\unit{s}{-1}\,\unit{Mpc}{-1}$, we get:
\be
\rho_{\tty{stars}}=7.1\cdot 10^{8}\ \Msol\,\unit{Mpc}{-3}
\label{rho_sca_eds}
\ee
which is about twice the value for the Salpeter IMF. These results are only very
weakly dependent on the cosmological model. Indeed, if we consider a
$\Lambda$CDM cosmological model $(\Omega_{0}=0.3,\Omega_{\Lambda}=0.7)$ and
calculate $\rho_{\tty{stars}}$, we find a difference that is negligible ($\sim
0.2$\,\%) for both a Salpeter and a Scalo IMF compared to our previous values.

\subsection{Energy density of the stellar radiation field}

Having obtained the energy radiated by stars (see Table~\ref{tab_star}, column
2) and the star density for a Salpeter IMF \equ{rho_salp_eds} and for a Scalo
IMF \equ{rho_sca_eds}, we calculate the energy density due to stars as
\be
U_{\tty{stars}}=\rho_{\tty{stars}}\cdot\left(\frac{E_{\tty{rad}}}{\Msol}\right)_{\tty{stars}}\ \tr{ergs}\,\unit{Mpc}{-3}\,.
\label{denset}
\ee
The obtained values are reported in the fourth column of Table~\ref{tab_star}.
Since $U_{\tty{stars}}$ is proportional to $\rho_{\tty{stars}}$ and
$E_{\tty{rad}}(\tss{stars})$, we note that $U_{\tty{stars}}$ decreases when $Z$
increases following the $E_{\tty{rad}}(\tss{stars})$ behavior and increases when
the IMF slope is steeper.

\section{Energy radiated by active galactic nuclei}\label{EnAGNs}

To estimate the energy radiated by AGN, we use the observed X-ray background
(XRB). This is motivated by the growing evidence that the XRB is emitted by
discrete sources that are mainly AGN as proposed by \citet{SetWol89}.
By adding the contribution of type I and II AGN, the observed XRB spectrum can
indeed be well reproduced \citep{Comastrial95, Gillial99, Gillial01}.
Furthermore, it seems that the contribution of both massive X-ray binaries and
supernovae to the XRB is negligible \citep{NatAlm00}. The emission from the hot
interstellar medium is also small compared to the AGN contribution
\citep{Comastrial95}. It is therefore possible to calculate the energy density
in the AGN radiation field, $U_{\tty{AGN}}$, from the observed XRB spectrum.
\citet{FabIwa99} give the following analytic parametrisation that describes well
the spectral energy distribution of the XRB as observed by HEAO-2 and ASCA for
soft and hard X-rays:
\be
I(E)=9\cdot E^{-0.4}\cdot \exp(-E/50\,\tty{keV})\ \tr{keV}\,\unit{cm}{-2}\,\unit{s}{-1}\,\unit{sr}{-1}\,\unit{keV}{-1}.
\label{paramXRB}
\ee
To relate this spectrum to the bolometric intensity of AGN, we use the mean
ratio from the $2-10$ keV intensity to the bolometric intensity derived by
\citet{FabIwa99}:
\be
\Lhx/\Lb\,\tr{(RQQ)}=3.3\,\%.
\label{fracRQQ}
\ee
This ratio is based on a compilation of spectral energy distributions of
radio-quiet quasars (RQQ) by \citet{Elvisal94}, which can be considered to be
representative for the average emission of unobscured AGN. We further assume
\citet{FabIwa99}'s hypothesis that the unobscured emission $E\,I(E)$ of AGN
down to 2 keV can be considered as constant with a value of
$38\,\tr{keV}\unit{cm}{-2}\unit{s}{-1}\unit{sr}{-1}$ which is the XRB intensity
at 30 keV. We then calculate the $2-10$ keV intensity of AGN as being
$61\,\tr{keV}\,\unit{cm}{-2}\,\unit{s}{-1}\,\unit{sr}{-1}$.
From this last value and Eq.~\eqp{fracRQQ}, we derive the bolometric intensity
of AGN and we calculate the energy density of the AGN radiation field as
\be
U_{\tty{AGN}}=3.7\,10^{58}\ \tr{ergs}\,\unit{Mpc}{-3}.
\label{densagn}
\ee

\section{Discussion}\label{discussion}

We have independently determined the energy densities emitted both by stars and
AGN in the universe. The obtained ratio $U_{\tty{AGN}}$/$U_{\tty{stars}}$ is
given in the last column of Table~\ref{tab_star} for two IMF and four different
metallicities. We note that all values are between $10^{-2}$ and $5\,10^{-2}$.
They slightly depend on the IMF slope and on the metallicity. Moreover, the
choice of the cosmological model has a negligible effect as mentioned at the end
of section \ref{evolStars}. This ratio is expected to remain rather constant
with cosmic time because the star formation history is similar to the AGN
luminosity evolution towards higher redshifts
\citep{Dunlop98,Franceschinial99}.

\subsection{Comparison between the different approaches}

Our results differ by several orders of magnitude from other estimations of the
energy ratio radiated by AGN and by stars. The different results extend from
$10^{-5}$ \citep{Csier01} to $10^{-1}$
\citep{FabIwa99,Franceschinial99,Elvisal02} through $4\,10^{-2}$ (this work). 
We cannot directly compare these values, because each study is based on a
different approach to the problem; sometimes observational and sometimes
approximative with the mass to energy conversion derived according to an
efficiency $\varepsilon_{\tty{rad}}$ through
$E_{\tty{rad}}=\varepsilon_{\tty{rad}} Mc^{2}$.

\citet{FabIwa99} consider a typical galaxy and the contribution to the flux
radiated both by the stars and the black hole hosted in this galaxy. They
implicitly assume all galaxies with a black hole to host an AGN. However, the
presence of a black hole does not mean that this galaxy is active. Some galaxies
might have had an active phase, but are currently quiescent. Taking their ratio
is equivalent to considering that every galaxy is currently active and
contributes to the radiated energy of AGN. This leads to overestimate the energy
radiated by AGN and thus to an overestimation of the AGN-to-star radiation
ratio.
 
\citet{Franceschinial99} tackles the problem from the observational point of
view starting with the energy radiated by the stars and the AGN to find a link
between both phenomena. This approach is motivated by the apparent similarity
between the cosmic evolution of the SFR and the AGN activity. 
In \citet{Csier01}, the sample of these active galaxies is based only on a radio
survey. Therefore, this selection results in a sample of radio-loud galaxies
which are only a small fraction of the whole AGN population. This leads to an
underestimation of the accretion rate per unit of volume because there are many
radio-quiet quasars that contribute to the energy radiated by accretion
processes that are not taken into account. The use of the $0.5-2$ keV volume
emissivity by \citet{Franceschinial99} allows one to include all AGN because
they all radiate in the X-rays
\citep{Comastrial95,Gillial99,Miyajial00,Pompilioal00,Gillial01} while only
including a negligible contribution of other objects like star clusters or
massive X-ray binaries \citep{Gillial99,NatAlm00,Gillial01}.

\citet{Elvisal02} also consider the AGN emission based on the XRB. However, they
compare it to the total luminosity of the universe that is estimated from the
diffuse background from submillimeter to UV wavelengths rather than only the
stellar luminosity.

We conclude that the various studies mentioned above actually measure different
quantities. Therefore, it is not possible to compare directly the values
obtained by different groups.

\subsection{Comparison between the different parameter values}

In addition to the differences in the approaches and the measured quantities
pointed out above, various studies use different values for the same
parameters. The stellar efficiency $\varepsilon_{\tty{rad}}$ used by
\citet{FabIwa99} is 0.0006, but it is of 0.001 in \citet{Franceschinial99}.
Similarly, the ratio of the $2-10$ keV luminosity to the bolometric luminosity
of the AGN is used in \citet{FabIwa99} and in this work, while
\citet{Franceschinial99} consider instead the $0.5-2$ keV flux to derive a
bolometric luminosity. Using the spectrum of Eq.~\eqp{paramXRB}, we can convert
their bolometric correction to the one based on the 2-10\,keV flux.
Thus, we derive that their bolometric correction differs from the 3.3\,\% value
of \citet{FabIwa99} by a factor of 3.
It seems therefore that there is an accumulation of different factors explaining
the diverging results in the literature. By taking the values of
$\varepsilon_{\tty{rad}}=0.0006$ and $\Lhx$/$\Lb=3.3\%$
from \citet{FabIwa99} and repeating the calculation of \citet{Franceschinial99},
we obtain a value of $\sim 0.053$ for the ratio
$\Lb\tss{(AGN)}$/$\Lb\tss{(SFR)}$, which tends to the value of 0.04 we obtain in
this work for a solar metallicity and a Salpeter IMF.

In order to explicitly calculate $U_{\tty{stars}}$ as defined here from the
study of \citet{Elvisal02}, we first subtract the quasar contribution to the
total luminosity of the universe they give to get only the stellar background.
We then obtain that $6.6\,10^{59}\ \tr{ergs}\,\unit{Mpc}{-3}$ are radiated by
stars which is quite similar to our result (see Table~\ref{tab_star}).
Therefore, keeping their value of $U_{\tty{AGN}}$, we estimate the ratio
$U_{\tty{AGN}}/U_{\tty{stars}}\sim 0.067$ that becomes $\sim 0.046$ if we use
the energy radiated by stars obtained in our study for a Salpeter IMF and a
solar metallicity \tab{tab_star} instead of theirs. If we compare this to our
results, we note that a difference also resides in the value of
$U_{\tty{AGN}}$. Considering \citet{Elvisal02}'s values of 48
$\tr{keV}\,\unit{cm}{-2}\,\unit{s}{-1}\,\unit{sr}{-1}$ at 30 keV for the XRB
instead of 38 $\tr{keV}\,\unit{cm}{-2}\,\unit{s}{-1}\, \unit{sr}{-1}$ and their
correction of a factor 1.6 to the bolometric correction, we recalculate our
value of $U_{\tty{AGN}}$ applying the same method as seen previously in
Sect.~\ref{EnAGNs}. We find that the AGN emitted energy density of
\citet{Elvisal02} is 1.5 times higher than ours. With this new derivation of
$U_{\tty{AGN}}$, we calculate a ratio $U_{\tty{AGN}}/U_{\tty{stars}}\sim 0.031$
for a Salpeter IMF and a solar metallicity that is a factor 2 lower than their
value of 0.067. The difference comes from a lower $U_{\tty{AGN}}$ and a higher
$U_{\tty{stars}}$, both effects combining to give a ratio twice lower.

Finally, it is worth noting that recent observations by \citet{Sarzial01} lead
to a smaller ratio of the central black hole mass to the host bulge mass than
the value used by \citet{FabIwa99}. By using their new result of
$M_{\tty{BH}}$/$M_{\tty{bulge}} \simeq 0.002$ the ratio
$E_{\tty{AGN}}$/$E_{\tty{stars}}$ found by \citet{FabIwa99} would have been of
$\sim\!0.07$ instead of $\sim\!0.2$.

\subsection{Relation between RLQ radio luminosity and AGN bolometric luminosity}\label{ratioLrToLb}

\citet{Franceschinial99} and \citet{Csier01} compared the star formation rate to
the accretion history based on the observation either in the X-ray band or in
the radio band. Every AGN is responsible for the X-ray emission but only a
subset of AGN has a significant radio emission. Therefore, we easily get a link
between these two classes of objects. 
When $1\ \Msol\,\unit{yr}{-1}\,\unit{Mpc}{-3}$ is used by the star formation,
$2.4\,10^{40}\ \tr{ergs}\,\unit{s}{-1}\,\unit{Mpc}{-3}$ are emitted by the AGN
in the $0.5-2$ keV band and $10^{-7}\ \Msol\,\unit{yr}{-1}\,\unit{Mpc}{-3}$ is
accreted onto the central black hole of a RLQ. If we transform the $0.5-2$ keV
luminosity into the bolometric luminosity using $\Lx$/$\Lb(\tr{AGN})\sim
1.25$\,\% derived from Eqs \eqpd{paramXRB}{fracRQQ} and if we consider an
accretion efficiency onto the black hole of 10\,\%, we obtain that
$3.4\,10^{-4}\ \Msol\,\unit{yr}{-1}\,\unit{Mpc}{-3}$ is accreted by the AGN for
a SFR of $1\ \Msol\,\unit{yr}{-1}\,\unit{Mpc}{-3}$. Therefore, we can compare
the luminosity of both the RLQ and the AGN as we get $\sim 10^{4}$ of difference
between them. When $1\ \tr{erg}\,\unit{s}{-1}\,\unit{Mpc}{-3}$ is radiated by
the RLQ at 2.7 GHz, the complete population of AGN radiates about 3400 times
more.

Furthermore, we can reconsider the method used by \citet{Csier01} to determine
$U_{\tty{AGN}}/U_{\tty{stars}}$. We have seen that we cannot directly compare
his result to the others since they are not considering the same family of
objects. Instead of using the analysis of RLQ by \citet{Dunlop98}, we
recalculate the ratio of \citet{Csier01} using the analysis of the XRB of
\citet{Franceschinial99}. Therefore, we obtain a value of
$E_{\tty{AGN}}$/$E_{\tty{stars}}\sim 0.015$ between the energies radiated by AGN
and stars instead of RLQ and stars. This value becomes $\sim 0.020$ if we use
our result of the energy radiated by stars for a Salpeter IMF and a solar
metallicity \tab{tab_star}. The initial inconsistency of a factor of a thousand
has been reduced to only a factor of 2 between this last estimate and the one in
Table~\ref{tab_star}.

\subsection{Mass accreted by supermassive black holes}

Based on our previous results we can estimate the total mass accreted by SMBH.
If we consider a galaxy with a mass of $10^{11}\ \Msol$ in stars, we can
estimate the energy radiated by those stars from the value in
Table~\ref{tab_star} for a Salpeter IMF and a solar metallicity. Using the
corresponding ratio of the AGN-to-stars energy release, we can then derive the
energy radiated by the SMBH. By further assuming an accretion efficiency of
10\%, we obtain that the total mass accreted by the SMBH is $6.5\,10^{7}\
\Msol$. In general, the typical mass accreted by a SMBH in a galaxy can be
calculated according to
\begin{equation}
\begin{split}
M_{\tty{AGN}}^{\tty{accr}}(\Msol) & =2.2\cdot 10^{7}\,
\left(\frac{0.1}{\varepsilon}\right)\!
\left(\frac{\scriptstyle E_{\tty{AGN}}/E_{\tty{stars}}}{0.04}\right)\! \\
& \times
\left(\frac{M_{\tty{Gal}}}{10^{11}\Msol}\right)\!
\left(\frac{(E_{\tty{rad}}/\Msol)_{\tty{stars}}}{10^{51}\ \tr{ergs}\,\Msol^{-1}}\right)\,.
\end{split}
\label{M_accr_agn}
\end{equation}

We can also estimate this mass starting from the link given in the previous
subsection. We have seen that when 1 $\Msol$ of material is converted into
stars, $3.4\,10^{-4}\ \Msol$ is accreted onto SMBH. If we consider again a
galaxy with a mass of $10^{11}\ \Msol$ in stars, we estimate that 
$3.4\,10^{7}\ \Msol$ is accreted onto the SMBH that is half of our previous
estimate. This difference by a factor of 2 is the same as for the ratio
$E_{\tty{AGN}}$/$E_{\tty{stars}}$ mentioned in Sect.~\ref{ratioLrToLb}.

\section{Conclusion}\label{conclusion}

We derived the relative contribution of AGN and stars to the radiation energy
density of the universe. The results are given for different IMF and
metallicities of the stellar population. The obtained values for the energy
ratio released by AGN and stars are all between 0.01 and 0.05. In the case of a
Salpeter IMF and solar metallicity we obtain a ratio of 0.04. This result cannot
be compared directly with previous studies because the approaches as well as the
values of some parameters used in the calculation differ from one study to the
other. Actually, when using similar parameters and appropriate correction
factors it seems that all previous studies do converge to values between 0.02
and 0.07 for the ratio of energy radiated by AGN and stars.

Since the CMB energy density is about $\sim 0.2\ \tr{eV}\,\unit{cm}{-3}$,
the energy density due to stars is of the order of 
$10^{60}\ \tr{ergs}\,\unit{Mpc}{-3}\sim 0.02\ \tr{eV}\,\unit{cm}{-3}$ 
and the AGN related radiation energy density is of about 
$8\,10^{-4}\ \tr{eV}\,\unit{cm}{-3}$, the general picture resulting from this
work is that the CMB contributes 10 times more than stars and 250 times more
than AGN to the local energy density of background photons.

We also estimate that RLQ contribute about 3400 times less than the whole AGN
family to the total accretion power in the universe and that the cumulated mass
accreted on average by a SMBH is of about $6.5\,10^{7}\,\Msol$.

\begin{acknowledgement}
We thank G. Meynet for useful discussions on star formation related issues.
\end{acknowledgement}

\bibliographystyle{aa}
\bibliography{BiblioArticle.bib}

\end{document}